\UseRawInputEncoding
\documentclass[floatfix,
superscriptaddress,
bibnotes,
 amsmath,amssymb,
 aps,
pra,
twocolumn
]{revtex4-2}
\usepackage{graphicx}
\usepackage{dcolumn}
\usepackage{xr}
\usepackage{xr-hyper}
\usepackage{hyperref}
\usepackage{bm} 
\usepackage{float}
\usepackage{hyperref} 
\hypersetup{
	colorlinks=true,
	linkcolor=blue,
	citecolor=blue,
	filecolor=magenta,
	urlcolor=blue         
}
\usepackage{color,soul}
\usepackage{lineno} 
\renewcommand{\selectlanguage}[1]{}

\begin{document}

\title{Nonequilibrium thermodynamics of the acoustoelectric quantum vacuum}

\author{Ryan O. Behunin}
\email[]{ryan.behunin@nau.edu}
\affiliation{Department of Physics, Northern Arizona University, Flagstaff, AZ 86011, USA}
\affiliation{Center for Materials Interfaces in Research and Applications (<MIRA!), Flagstaff, AZ, USA}

\author{Andrew J. Shepherd}
\affiliation{Department of Physics, Northern Arizona University, Flagstaff, AZ 86011, USA}
\affiliation{Center for Materials Interfaces in Research and Applications (<MIRA!), Flagstaff, AZ, USA}

\author{Francesco Intravaia}
\affiliation{
Humboldt-Universit\"at zu Berlin, Institut f\"ur Physik, 12489 Berlin, Germany
}

\author{Matt Eichenfield}
\affiliation{Electrical, Computer, and Energy Engineering, University of Colorado Boulder and Sandia National Laboratories}

\date{\today}

\begin{abstract}
The quantum vacuum can assume thermal properties as a consequence of system kinematics, highlighting the nuance of our definition of particles in quantum field theory. Here, we explore this phenomenon in acoustoelectric systems, involving the interaction of phonons and plasmons, where the charge carriers drift at a constant velocity exceeding the speed of sound. Through an open quantum systems analysis, we show that the acoustoelectric quantum vacuum acquires a thermal character with a temperature defined by the drift velocity and the phonon wavevector. Realistic parameters yield effective temperatures of several Kelvin, establishing acoustoelectric systems as a promising platform for the investigation of quantum vacuum effects.
\end{abstract}

\maketitle

One of the most striking results of quantum field theory is that the presence or absence of particles can depend on the kinematics of a system \cite{Fulling1973,Davies1975,Unruh1976,birrell1984quantum,milonni2013quantum}. The Unruh effect is a prime example of this physics, where the quantum vacuum (i.e., the state devoid of particles) is perceived by a non-inertial observer of uniform proper acceleration $a$ to be populated with an ensemble of particles in a thermal state of temperature
\begin{align}
    T_U = \frac{\hbar a}{2\pi k_B c}.
\end{align}
Recent work has further elucidated this effect, showing that the `perception' of the vacuum will be altered when a change of frame or configuration leads to an anomalous Doppler effect, resulting in negative frequencies \cite{svidzinsky2019excitation,svidzinsky2021unruh}. The physical origin of these negative frequencies can be understood by considering an observer moving parallel to a wave of frequency $\omega$ and wavevector ${\bf k}$. As the observer's velocity ${\bf v}$ increases from below to above the wave's phase velocity, the Doppler-shifted frequency $\omega' = \omega-{\bf v}\cdot {\bf k}$ changes sign, and the wave appears to reverse its direction of propagation. Interestingly, modes with negative frequency carry negative energy, so that their excitation lowers the total energy of the system \cite{nezlin1976negative}. In the Fulling--Davies--Unruh effect, such negative-frequency modes arise as a consequence of non-inertial motion \cite{svidzinsky2021unruh}. Remarkably, in systems possessing a preferred reference frame, such as those containing matter, negative frequencies can also arise for inertial observers, where the vacuum defined with respect to the laboratory frame is perceived to be approximately thermal \cite{intravaia2016vacuum,reiche2026nonequilibrium}.

\begin{figure}
    \centering
    \includegraphics[width=0.99\linewidth]{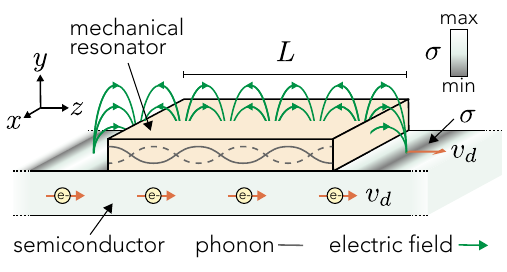}
    \caption{Mechanical resonator coupled electromechanically to moving charge oscillations. When the free carriers are set in motion with drift velocity $v_d$, fluctuating surface charge density $\sigma$ sources evanescent electric fields that can drive dynamic mechanical displacement.}
    \label{fig: system diagram}
\end{figure}

Here, we explore the nature of the quantum vacuum in acoustoelectric systems of the type shown in Fig. \ref{fig: system diagram}. We analyze the quantum dynamics of a mechanical oscillator that is electromechanically coupled to a semiconducting medium containing a constant drift current (see Refs. \cite{shapiro2010thermal,shapiro2017fluctuation} for related examples). This acoustoelectric system permits complex energy transfer between the mechanical motion and the oscillations of the moving free carriers. 
Working from a Hamiltonian framework, we derive the Heisenberg-Langevin equation for the oscillator, capturing the backreaction and the noise imparted by the free-carriers \cite{behunin2026noise}. This formalism permits the calculation of nonequilibrium power spectra for the oscillator variables that we analyze within the framework of the fluctuation-dissipation relation \cite{callen1951irreversibility}. In the presence of a steady drift current, this analysis leads to drift-velocity dependent fluctuations that are well-described by a thermal equilibrium power spectrum. This is made possible because a subset of the plasmons undergoes an anomalous Doppler shift; the realization of these negative (lab frame) frequencies permits the excitation of the mechanical resonator by the emission of a plasmon. This process populates the oscillator, driving it into an effective thermal state with a temperature characterized by the drift velocity $v_d$ and the spatial frequency $q$ of the mechanical mode  
\begin{align}
\label{eq: effective temperature}
    T_{AE} \approx \frac{\hbar v_d q}{2 k_B}.
\end{align}
Much like atoms in relative motion to matter \cite{intravaia2016vacuum,svidzinsky2019excitation,reiche2026nonequilibrium}, this result shows that the nonequilibrium quantum vacuum fluctuations of acoustoelectric systems behave similarly to systems coupled to a heat bath.
Remarkably, such temperatures may reach detectable levels. 
For experimentally achievable drift velocities and phonon spatial frequencies, $v_d \sim 50,000$ m/s and $2\pi/q \sim 500$ nm, the effective temperature of the mechanical mode can reach $T_{AE} \sim 2.4$K. These results highlight the potential for acoustoelectric devices to serve as a testbed for nonequilibrium features of the quantum vacuum.  

{\it System model and Hamiltonian:} The acoustoelectric effect can take place within systems that possess free carriers and some form of electromechanical coupling, e.g., piezoelectricity \cite{Parmenter}. In such systems, electric fields can be produced by oscillations of free charge, i.e., plasmons, that can generate mechanical strain. By applying a bias voltage, the free-carriers can be set in motion with drift velocity ${\bf v}_d$, permitting energy transfer between mechanical vibrations and charge motion \cite{Hutson1961CdS,collins1968amplification,coldren1971amp,coldren1973cw}.

Acoustoelectric interactions mediate two phonon-plasmon scattering processes that can conserve energy and momentum \cite{behunin2026noise}. In the first process, a phonon of respective angular frequency and wavevector $(\Omega,{\bf q})$ can be annihilated and converted to a plasmon of lab frame frequency $\omega' = \Omega$ and wavevector ${\bf k} = {\bf q}$, leading to mechanical dissipation. Given that the plasmon frequency in the frame of the drift current $\omega = \omega'-{\bf v}_d\cdot{\bf k}$ must be positive, energy and momentum conservation require $ \omega = \Omega(1-v_d/v_m\cos \phi)>0$. Here, $\phi$ is the angle between the phonon wavevector and the drift current and $v_m$ is the magnitude of phase velocity of the mechanical wave (i.e., $\Omega = v_m |{\bf q}|$). These conditions show that forward scattering is permitted for all $\phi$ such that $v_d \cos \phi < v_m$. In the second process, a phonon and plasmon of respective angular frequency and wavevector $(\Omega,{\bf q})$ and $(\omega',{\bf k})$ are spontaneously emitted from the vacuum. Although this results in the apparent creation of two particles from nothing, this process can conserve energy and momentum, i.e., ${\bf \omega}' + \Omega =0$ and ${\bf k} + {\bf q} =0$, if the plasmon propagates anti-parallel to the phonon and lab frame frequency of the plasmon is {\it negative}. By requiring $\omega = \omega'-{\bf v}_d\cdot{\bf k}>0$, we find $v_d/v_m\cos \phi-1>0$, defining, in a manner analogous to the Cherenkov effect, a cone of angle $\phi_c = \cos^{-1}(v_m/v_d)$ wherein this scattering process is permitted. This scattering process allows the mechanical resonator to be excited through the emission of a negative frequency plasmon.

The acoustoelectric system considered here is comprised of a mechanical resonator of length $L$ that is electromechanically coupled to moving charge oscillations (Fig. \ref{fig: system diagram}) that are set in motion by a constant drift current \cite{wendt2026electrically}. 
We narrow our focus to a single standing-wave mechanical mode of spatial frequency $q$ along the $z$-axis, and to a set of traveling-wave surface plasmon modes that co- and counter-propagate with the drift current. We assume that $q L \gg 1$ (i.e., we consider a high overtone of the mechanical resonator), enabling traveling-wave plasmons of spatial frequency $\pm q$ to accurately capture the acoustoelectric dynamics.  

The total system Hamiltonian $H = H_0 + H_{int}$ is comprised of a component representing the free evolution of the phonon mode and the plasmon bath,
\begin{align}
\label{eq: H_0}
    {H_0} = \hbar \Omega b^\dag b +\sum_{j=\pm } \int_0^\infty \!\!\!\! d\omega \ \hbar  \omega_j a^\dag_j(\omega) a_j(\omega), 
\end{align}
where $j = + (-)$ labels the plasmon mode that copropagates (counterpropagates) with the drift current, and $\omega_j = \omega +j v_d q$ is the Doppler-shifted plasmon frequency observed in the lab frame.
The interaction Hamiltonian $H_{int}$ quantifies the acoustoelectric coupling 
\begin{align}
\label{eq: H_int}
    H_{int} =  \sum_{j=\pm } \int_0^\infty \!\! \!\! d\omega \ \hbar g\left(\chi_\omega a_j(\omega)+\chi^*_\omega a^\dag_j(\omega)\right) \left(b+b^\dag\right).
\end{align}
Here, the drift velocity is assumed to be held constant by an applied electric field, and $a_{\pm}(\omega)$ ($a^\dag_{\pm}(\omega)$) and $b$ ($b^\dag$) respectively represent the annihilation (creation) operators for the plasmon modes of angular frequency $\omega$ (defined in the frame moving at $v_d$) and wavevector $\pm q$, as well as the phonon mode \cite{behunin2026noise}. These operators satisfy commutation relations
\begin{align}
       & [a_j(\omega),a^\dag_{j'}(\omega')] = \delta_{jj'}\delta(\omega - \omega')
       \\
       &  [a_j(\omega),a_{j'}(\omega')] = [a^\dag_j(\omega),a^\dag_{j'}(\omega')] = 0 
      \\
      & [b,b^\dag] = 1
      \\
      &  [b,b] = [b^\dag,b^\dag] = 0.
    \end{align}
The acoustoelectric coupling is denoted by $g$, and the susceptibility function $\chi_\omega$ describes damped plasma oscillations
\begin{align}
    \chi_\omega = \sqrt{\frac{2\gamma \omega_0}{\pi }}\frac{\sqrt{\omega}}{-\omega^2 - i \gamma \omega + \omega_0^2},
\end{align}
where $\gamma$ is the decay rate and $\omega_0$ is the resonant frequency of the plasmons \cite{barton1979some,barton1997van,behunin2026noise}. 
This model assumes the validity of the quasielectrostatic approximation, i.e., $\nabla \times {\bf E} = 0$ where ${\bf E}$ is the electric field,  and reproduces the plasmon dynamics linearized about the uniform drift velocity $v_d$ \cite{behunin2026noise}.   

{\it Heisenberg-Langevin equations:}
Information regarding the nonequilibrium steady-state of this system can be calculated using the Heisenberg-Langevin equations. Beginning from the Heisenberg equations of motion for the annihilation operators given by
\begin{align}
\label{eq: plasmon HOM}
   & \dot{a}_j(\omega) =  -i \omega_j a_j(\omega) - i g \chi^*_\omega (b+b^\dag),
    \\
    \label{eq: phonon HOM}
    & \dot{b} =  - i\Omega b - i \sum_{j=\pm} \int_0^\infty \!\! \!\! d\omega \ g(\chi_\omega a_j(\omega)+\chi^*_\omega a^\dag_j(\omega)),
\end{align}
the formal solution for the plasmon operators can be obtained and inserted into Eq. \eqref{eq: phonon HOM} to describe the effective dynamics of $b$. After some simplification, this leads to the Heisenberg-Langevin equation \cite{ford1988quantum,breuer2002theory} for the oscillator position operator $x = \sqrt{\frac{\hbar}{2 \Omega}}(b+b^\dag)$  given by
\begin{align}
\label{eq: Heisenberg-Langevin}
\ddot{x}(t) + \Omega^2 x(t) =  \int_{-\infty}^\infty \!\!\!\! d\tau \ \mu(\tau) x(t-\tau) + \xi(t),
\end{align}
where $\mu(\tau)$ captures the mechanical dissipation (and amplification) and dispersion produced by the plasmon bath
\begin{align}
\label{eq: susceptibility}
    \mu(\tau) = 4 \Omega \sum_{j=\pm }\int_0^\infty \!\!\!\! d\omega \  |g|^2|\chi_\omega|^2
    \theta(\tau) \sin (\omega_j \tau). 
\end{align}
In Eq. \eqref{eq: susceptibility}, $\theta(\tau)$ is the Heaviside $\theta$-function, and $\xi$ is the Langevin force describing how plasmon fluctuations impart noise on the phonon mode
\begin{align}
    \xi(t) = - \sqrt{2 \hbar \Omega}  \sum_{j=\pm } \int_0^\infty  d\omega \ g\bigg( & \chi_\omega \hat{a}_j(\omega)e^{-i\omega_j t} \nonumber
    \\ & \quad +\chi^*_\omega \hat{a}^\dag_j(\omega)e^{i \omega_j t}\bigg)  
\end{align}
(see Appendix \ref{Sec: Heisenberg-Langevin Derivation}). 
Here, $\hat{a}_j(\omega)(\hat{a}^{\dag}_j(\omega))$ denotes the Schr\"odinger picture annihilation (creation) operator for the plasmon bath (where we have reserved the $\ \hat{}\ $ symbol for operators describing the free-evolution of the plasmons). 

{\it Nonequilibrium quantum fluctuations}: 
The nonequilibrium quantum fluctuations of the phonon mode can be quantified by the power spectrum of  the position operator $S_x(\omega)$. According to the Wiener-Khinchin theorem,
\begin{align}
S_x(\omega) = \int_{-\infty}^\infty d\tau \ e^{ i\omega\tau} \langle x(t+\tau) x(t) \rangle
\end{align}
where $\langle ... \rangle$ denotes the expectation value with respect to the initial state $|\psi_i\rangle$ of the system.  Owing to the linearity of Eq. \eqref{eq: Heisenberg-Langevin}, the solution can be efficiently obtained in the Fourier domain, and expressed as
\begin{align}
\label{eq: solution x}
    x(t) =  \frac{1}{2\pi}\int_{-\infty}^\infty \!\!\!\! d\tau  \int_{-\infty}^\infty \!\! \!\! d\omega \ e^{-i \omega(t-\tau)} \alpha(\omega) \xi(\tau)
\end{align}
where $\alpha(\omega) = (-\omega^2+\Omega^2-\tilde{\mu}(\omega))^{-1}$ is the frequency domain representation of the oscillator's susceptibility function, i.e., the retarded Green's function for Eq. \eqref{eq: Heisenberg-Langevin}, 
and $\tilde{\mu}(\omega) = \int_{-\infty}^\infty d\tau \exp\{ i\omega\tau\} \mu(\tau)$ is
\begin{align}
\label{eq: mu-susc}
    \tilde{\mu}(\omega) 
= & 
\sum_j \int_0^\infty d\omega' \frac{4\Omega |g|^2\omega_j'}{{\omega_j'}^2-(\omega+i 0^+)^2}|\chi_{\omega'}|^2.
\end{align}

Assuming the plasmons are prepared in the quantum ground state (i.e.,  $\hat{a}_j(\omega)|\psi_i\rangle = 0$), the two-time correlation function for the Langevin force can be explicitly calculated
\begin{align}
\label{eq: Langevin correlation}
    \langle \xi(\tau)\xi(\tau')\rangle = 2 \hbar \Omega \sum_{j=\pm} \int_0^\infty \!\!\!\! d\omega |g|^2 |\chi_\omega|^2 e^{-i \omega_j(\tau-\tau')}
\end{align}
where $ \langle \hat{a}_j(\omega) \hat{a}^\dag_{j'}(\omega')\rangle = \delta_{jj'}\delta(\omega-\omega')$ has been used. The Fourier transform of Eq. \eqref{eq: Langevin correlation} gives the power spectrum $S_\xi(\omega)$ of the force fluctuations that drive the phonon mode
\begin{align}
\label{eq: xi-power spectrum}
    S_\xi(\omega) 
     = & 2\hbar (\Gamma_+ \theta(\omega_+)+ \Gamma_- \theta(\omega_-))
\end{align}
which, using Eq. \eqref{eq: solution x}, provides the oscillator power spectrum $S_x(\omega) = |\alpha(\omega)|^2 S_\xi(\omega)$. Here, we have introduced the notation $\Gamma_j = 2\pi \Omega|g|^2 |\chi_{\omega_j}|^2$, parameters that quantify the energy transfer between the oscillator and plasmons of lab frame frequency $\omega_j$.  

Next, we leverage the analytical properties of $\alpha(\omega)$ and $\tilde{\mu}(\omega)$ to relate the oscillator power spectrum to the mechanical dissipation. First, we multiply the equation for the susceptibility $(-\omega^2+\Omega^2 - \tilde{\mu}(\omega))\alpha(\omega) = 1$ by $\alpha^*(\omega)$ and take the imaginary part to obtain 
\begin{align}
   {\rm Im}[\alpha(\omega)] = |\alpha(\omega)|^2 {\rm Im}[\tilde{\mu}(\omega)].
\end{align}
The imaginary part of $\alpha(\omega)$ quantifies the mechanical dissipation rate and explicit calculation of $\tilde{\mu}(\omega)$ using Eq. \eqref{eq: mu-susc} gives
\begin{align}
    {\rm Im}[\tilde{\mu}(\omega)] = \Gamma_+ + \Gamma_-. 
\end{align}
\begin{figure}
    \centering
    \includegraphics[width=\linewidth]{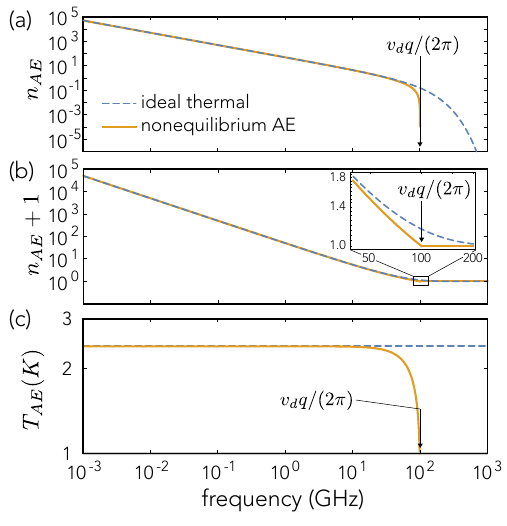}
    \caption{Nonequilibrium occupation number (a) and (b), and effective temperature (c). Results are compared to thermal equilibrium results with $T = \hbar v_d q/2k_B$. The parameters used are $v_d = 50,000$ m/s, $q = (2\pi)2 \times 10^6$ 1/m, $\omega_0 = (2\pi) 5$ THz, and $\gamma = (2\pi) 5.1$ THz. Note that the plotted quantities are independent of the coupling $g$.}
    \label{fig: nonequilibrium AE results}
    \end{figure}

\noindent 
Combining these results leads to the nonequilibrium form of the fluctuation-dissipation relation given by 
\begin{align}
\label{eq: non-eq FDR}
S_x(\omega) = 2 \hbar 
 \frac{ 
\Gamma_+ \theta(\omega_+)+ \Gamma_- \theta(\omega_-)}{\Gamma_+ + \Gamma_-} 
  {\rm Im}[\alpha(\omega)].
\end{align}
Note two things: (1) in the limit that $v_d \to 0$, Eq. \eqref{eq: non-eq FDR} correctly reproduces the equilibrium fluctuation-dissipation relation at zero temperature $S_x(\omega) = 2 \hbar \theta(\omega ){\rm Im}[\alpha(\omega)]$, and (2) Eq. \eqref{eq: non-eq FDR} takes the same form as the thermal equilibrium fluctuation-dissipation relation, i.e., $S_x(\omega) = 2 \hbar 
 (n_{AE}(\omega)+1) 
  {\rm Im}[\alpha(\omega)]$,  where the effective thermal occupation $n_{AE}(\omega)$ is given by 
\begin{align}
\label{eq: effective thermal occupation}
  n_{AE}(\omega)= 
       &   -\frac{ 
\Gamma_+ \theta(-\omega_+)+ \Gamma_- \theta(-\omega_-) }{\Gamma_+ + \Gamma_-} 
\end{align}
(note that for $\omega_- <0$ that $\Gamma_- <0$).
This result and dimensional analysis suggest that the effective temperature follows the relation, $T_{AE} \sim \hbar v_d q/k_B$, as we expect $n_{AE} \ll 1$ for $\hbar \omega \gg k_B T_{AE}$. By assuming $n_{AE}(\omega)$ takes the form of a Boltzmann distribution, Eq. \eqref{eq: effective thermal occupation} can be formally solved for the effective temperature 
\begin{align}
\label{eq: freq-dep temp}
    T_{AE}(\omega) = 
    \left\{\begin{array}{cc}
            - \frac{\hbar \omega/k_B}{\ln (-\Gamma_+/\Gamma_-)} & \quad {\rm for } \quad v_d q > \omega > 0
         \\
          0 & \quad {\rm for }   \quad \omega > v_d q > 0.
    \end{array} \right.
\end{align}
In the limit where $\gamma >\omega_0 \gg v_d q \gg \omega$, Eq. \eqref{eq: freq-dep temp} can be expanded for small $\omega$, leading to the frequency-independent temperature given by Eq. \eqref{eq: effective temperature}. See Appendix \ref{Sec: NEq-FDR} for an alternative derivation of this result utilizing symmetries of the two-time correlation functions.  
\begin{figure}[t]
    \centering
    \includegraphics[width=\linewidth]{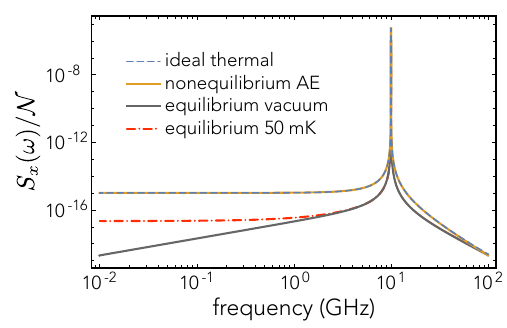}
    \caption{Normalized oscillator power spectrum for a nonequilibrium steady-state (orange), thermal state with $T=\hbar v_d q/(2kB)$ (blue dashed),  equilibrium state at zero temperature (gray), and thermal state with $T = 50$ mK (red, dot-dashed). The parameters used are the same as Fig. \ref{fig: nonequilibrium AE results} in addition to $g = 2$ GHz (see Ref. \cite{behunin2026noise}), $\Omega = (2\pi) 10$ GHz, and $v_m$ = 5000 m/s. Each power spectrum is normalized by $\mathcal{N} = S_x(\Omega)$ using Eq. \eqref{eq: non-eq FDR}.}
    \label{fig:power spectrum}
\end{figure}

Figure \ref{fig: nonequilibrium AE results} compares the effective temperature and phonon occupation number with the corresponding thermal equilibrium results at temperature $T = \hbar v_d q/(2 k_B)$. Using realistic values for the parameters (see caption), we find that the nonequilibrium fluctuations are well-described by a thermal state for $\omega < v_d q$. In this regime, a subset of the plasmon modes can experience an anomalous Doppler shift, which enables spontaneous emission of phonon-plasmon pairs from vacuum. Because the oscillator samples these fluctuations in a narrow band centered at $\Omega$, measurements of the oscillator power spectrum will (1) be virtually indistinguishable from the analogous thermal equilibrium result (see Appendix Sec. \ref{Sec: NEq-FDR}) when $v_dq \gg \Omega$ as shown in Fig. \ref{fig:power spectrum}, and (2) be distinct from the zero-temperature equilibrium power spectrum. Put together, these results show that when quantum fluctuations of plasmons are Doppler-shifted by carrier drift, they take on a thermal character similar to the exponential redshifting of vacuum fluctuations that occurs in the Fulling-Davies-Unruh effect \cite{birrell1984quantum}.  

Figure \ref{fig:power spectrum} also shows the impact of Joule heating on these results. By computing the power dissipated by the steady-drift current, the system base temperature can be estimated using typical cryostat cooling powers. For a device with a $300$ nm wide, $150$ nm thick and $100 \mu$m long semiconducting region with a carrier concentration  $N = 10^{16}$ (cm)$^{-3}$ and mobility of $2000$ cm$^2$/Vs, the Joule heating is $\sim 90 \ \mu$W for $v_d = 50,000$ m/s. This heat load can be balanced by the cooling power in standard dilution refrigerators at a temperature $~50$ mK. So long as $T_{AE} > T$, such heating leads to a perturbative correction to the nonequilibrium power spectrum given by Eq. \eqref{eq: non-eq FDR} (see Appendix \ref{Sec. Heating}).  Under these conditions, Fig. \ref{fig:power spectrum} shows that the nonequilibrium vacuum fluctuations dominate any thermal fluctuations produced by heating.

{\it Discussion:}
The investigation of the quantum vacuum has enriched our understanding of quantum field theory, elucidated the very nature of what constitutes a particle, and revealed that black holes evaporate \cite{Fulling1973,Hawking1974,Davies1975,Unruh1976}.  
We have shown, by analogy with the equilibrium fluctuation-dissipation theorem, that an effectively thermal state emerges from the quantum vacuum of acoustoelectric systems when the free carriers move with a constant velocity. 
In contrast with traditional systems requiring relativistic velocities or accelerations, the relevant velocity scale of acoustoelectric systems is set by the speed of sound, drastically reducing the challenges to observing these effects. 
Emerging acoustoelectric systems are poised to observe this physics, as they can achieve large drift velocities, be operated within dilution refrigerators, and the phonons can be sensitively probed \cite{wendt2026electrically}. 
These results highlight acoustoelectric systems as a promising testbed for quantum vacuum phenomena and nonequilibrium thermodynamics.

{\it Acknowledgments}---
This work was supported by NSF Awards No. 2145724 and No. 2427169.

\bibliography{refs}

\appendix

\section{Derivation of Heisenberg-Langevin equation}
\label{Sec: Heisenberg-Langevin Derivation}
In this section, we provide more details on the derivation of Eq. \eqref{eq: Heisenberg-Langevin}. Inserting the formal solution to Eq. \eqref{eq: plasmon HOM} given by
\begin{align}
    a_j(\omega,t) = & \hat{a}_j(\omega)e^{-i\omega_jt} 
    \\
    & 
     -ig \int_{-\infty}^t  d\tau \  \chi_\omega^* e^{-i\omega_j(t-\tau)}(b(\tau)+b^\dag(\tau)) \nonumber
\end{align}
into Eq. \eqref{eq: phonon HOM} we obtain
\begin{align}
\label{eq: b-Langevin}
    \dot{b}(t) =  -i\Omega b(t) & +  \frac{i}{\sqrt{2 \hbar \Omega}}\xi(t) 
    \\
     & +  \frac{i}{2\Omega}\int_{-\infty}^\infty d\tau \ \mu(t-\tau)(b(\tau) + b^\dag(\tau)) \nonumber
\end{align}
Adding and subtracting Eq. \eqref{eq: b-Langevin} from its Hermitian conjugate (and defining the quadrature operators $X = b+b^\dag$ and $Y = -i(b-b^\dag)$) gives
\begin{align}
    & \dot{X}(t) = \Omega Y(t)
    \\
    & \dot{Y}(t) = -\Omega X(t)
    + \sqrt{\frac{2}{\hbar \Omega}}\xi(t)
    + \frac{1}{\Omega}\int_{-\infty}^\infty d\tau \ \mu(t-\tau) X(\tau).
\end{align}
Eliminating $Y$ and scaling by $\sqrt{\hbar/2\Omega}$ leads to Eq. \eqref{eq: Heisenberg-Langevin}. 

\section{Finite temperature nonequilibrium fluctuation theorem}
\label{Sec: NEq-FDR}
At finite temperature (defined in the lab frame), the oscillator power spectrum is given by 
\begin{align}
\label{Eq: thermal equ. power spectrum}
    S_x(\omega) = \int_{-\infty}^\infty d\tau \ {\rm tr}\{\rho_{eq} x(t+\tau)x(t)\}.
\end{align}
Here, $\rho_{eq}$ is the density matrix for the system in thermal equilibrium (i.e., $v_d =0$)
\begin{align}
    \rho_{eq} = \frac{1}{Z} e^{-\beta H_{eq}},
\end{align}
with inverse temperature $\beta = 1/(k_B T)$, $H_{eq}$ given by
\begin{align}
    H_{eq} = \hbar \Omega b^\dag b +\sum_{j=\pm } \int_0^\infty \!\!\!\! d\omega \ \hbar   \omega a^\dag_j(\omega) a_j(\omega) + H_{int},
\end{align}
and where $Z$ is the partition function. Within the expectation value, $x(t)$ is the Heisenberg picture operator evaluated with respect to $H$ (i.e., the Hamiltonian accounting for charge motion with $v_d \neq 0$). 
This derivation relies on two results: (1) that the total Hamiltonian can be diagonalized (see Supplementary Information)
\begin{align}
    H = \sum_{j = \pm} \int_0^\infty d\omega \ \hbar \omega_j \hat{a}^\dag_j(\omega) \hat{a}_j(\omega) 
\end{align}
in terms of the Schr\"odinger picture plasmon operators $\hat{a}_j(\omega)$ and $\hat{a}^\dag_j(\omega)$ 
and (2) that $x$ can be expressed as $x = \sum_j x_j $ where 
\begin{align}
   x_j = -\sqrt{2\hbar \Omega} g \int_0^\infty d\omega \ [ &
   \alpha(\omega_j) \chi_\omega \hat{a}_j(\omega) 
   \\
   &+
   \alpha^*(\omega_{-j}) \chi^*_\omega \hat{a}^\dag_{-j}(\omega)]. \nonumber
\end{align}

Using these results, we can derive symmetry properties of the two-time correlation functions. Noting that ${\rm tr}\{\rho_{eq} x_j(t+\tau)x_{j}(t)\} =0$, we find ${\rm tr}\{\rho_{eq} x(t+\tau)x(t)\} = \sum_j C_j(\tau)$ where 
\begin{align}
   C_j(\tau) = {\rm tr}\{\rho_{eq} x_j(t+\tau)x_{-j}(t)\}.
\end{align}
Using the cyclic property of the trace, $C_j(\tau)$ can be expressed as
\begin{align}
   C_j(\tau) = {\rm tr}\{\rho_{eq} x_{-j}(t) e^{-\beta H_{eq}} x_j(t+\tau) e^{\beta H_{eq}}\}.
\end{align}
After using the Baker-Campbell-Hausdorff formula
\begin{align}
    e^{-\beta H_{eq}} x_j(t+\tau) e^{\beta H_{eq}} = e^{\hbar \beta v_d q j} x_j(t+\tau + i \hbar \beta),
\end{align}
we obtain the following periodicity in imaginary time for two-time correlation functions in thermal equilibrium 
\begin{align}
\label{Eq: im-time periodicity}
   C_j(\tau) = e^{\hbar \beta v_d q j} C_{-j}(-\tau-i\hbar \beta). 
\end{align}
Defining the generalized susceptibility $\alpha(\tau) = \sum_j \alpha_j(\tau)$, we find 
\begin{align}
   \alpha_j(\tau) = & \frac{i}{\hbar}{\rm tr}\{\rho_{eq}[x_j(t+\tau),x_{-j}(t)] \}, 
   \\
   = & \frac{i}{\hbar} (C_j(\tau) -C_{-j}(-\tau)),
   \\
   = & \frac{i}{\hbar} (C_j(\tau) -e^{-\hbar \beta v_d q j}C_{j}(\tau-i\hbar \beta)),
\end{align}
where Eq. \eqref{Eq: im-time periodicity} has been used in the last line. Defining  $S_j(\omega) = \int_{-\infty}^\infty d\tau \exp\{i\omega 
\tau\}C_j(\tau)$ and using that $C^*_j(\tau) = C_j(-\tau)$, it can be shown that $S_j(\omega)$ is real-valued. Using these results, the imaginary part of the Fourier transform of $\alpha_j(\tau)$ is given by
\begin{align}
    {\rm Im}[\alpha_j(\omega)]  
     = &  \frac{1}{2\hbar}(1-e^{-\hbar \omega_j \beta})S_j(\omega).
\end{align}
Inserting this result into Eq. \eqref{Eq: thermal equ. power spectrum}, we obtain
\begin{align}
\label{Eq: Nonequilibrium FDR finite T}
    S_x(\omega,T) 
    = & 2\hbar \frac{\Gamma_+(n(\omega_+)+1)+ \Gamma_-(n(\omega_-)+1)}{\Gamma_+ + \Gamma_-} {\rm Im}[\alpha(\omega)],
    \end{align}
where $n(\omega) = (\exp\{\hbar \omega \beta\} -1)^{-1}$, and the properties of $\alpha(\omega)$ have been utilized. In the limit, that $T \to 0$ Eq. \eqref{Eq: Nonequilibrium FDR finite T} reduces to Eq. \eqref{eq: non-eq FDR}, providing an alternative derivation to the nonequilibrium fluctuation-dissipation relation, and the oscillator power spectrum in thermal equilibrium at temperature $T$ can be obtained by taking $v_d\to 0$ in Eq. \eqref{Eq: Nonequilibrium FDR finite T}
\begin{align}
\label{Eq: thermal equilibrium FDR}
    S_x(\omega) = 2\hbar (n(\omega)+1){\rm Im}[\alpha(\omega)],
\end{align}
where it is understood that $v_d \to 0$ must be taken in $\alpha(\omega)$ as well.

\section{Heating produced by drift current}
\label{Sec. Heating}
To investigate the experimental accessibility of these effects, we estimate the Ohmic heating produced by the drift current. Such heating has the potential to raise the system temperature, making it impossible to observe these quantum vacuum effects. First, we estimate the dissipated power density $p_d$ given by 
\begin{align}
    p_d = \frac{1}{\mu_e}{\bf J}\cdot{\bf v}_d
\end{align}
where $\mu_e$ is the free-carrier mobility and ${\bf J} = e N {\bf v}_d$ is the current density given in terms of electron charge $e$ and free-carrier density $N$.  Using realistic parameters for InGaAsP, $N = 10^{16} \ ({\rm cm})^{-3}$, $\mu_e = 2000 \ {\rm cm}^2/{\rm V s}$, and $v_d = 50,000 \ {\rm m/s}$, the dissipated power density is $p_d = 2 \times 10^7 {\rm W}/{\rm cm}^3$. Next, we specify the dimensions of the device to quantify the total dissipated power. Assuming that the semiconducting region is a thin strip 300 nm wide, 150 nm thick and 100 $\mu$m long, the total dissipated power is approximately $90 \ \mu$W, falling well below cooling powers ($\sim 400\ \mu$W) that can be achieved in dilution refrigerators at 100 mK---cold enough for $10$ GHz phonons to occupy the quantum ground state with probability surpassing $99\%$. Using the low-temperature dilution refrigerator cooling power $\dot{Q} = \dot{n}_3(a T_m^2-b T^2_{in})$, with $\dot{n}_3$ the He$_3$ molar circulation rate, $T_m$ the mixing plate temperature, $T_{in}$ the temperature of the He$_3$ gas entering the mixing chamber, and $a$ and $b$ empirical constants, standard $14 \ \mu$W cooling powers at $20$ mK can be extrapolated to intermediate temperatures, showing that a temperature of $50$ mK can be achieved with a $90 \ \mu$W heat load.  

Further refinements in device geometry, carrier concentration, and electron mobility have the potential to significantly reduce this dissipated power. For example, two-dimensional electron gases (2DEG) have mobilities surpassing $10^7 {\rm cm}^2/{\rm V s}$, and have (two-dimensional) carrier concentrations of $N_e \sim 10^{10} ({\rm cm})^{-2}$ \cite{kumar2010nonconventional,shtrikman1999n11}. With a 300 nm wide and 100 $\mu$m long 2DEG, the dissipated power is 1.2 nW. 

\subsection{Thermal corrections to Eq. \eqref{eq: non-eq FDR}}

Here, we estimate corrections to the nonequilibrium power spectrum produced by Joule heating. Using Eq. \eqref{Eq: Nonequilibrium FDR finite T}, the ratio of the nonequilibrium power spectra at zero- and finite-temperature gives
\begin{align}
    \frac{S_x(\omega,T)}{S_x(\omega)} = \frac{\Gamma_+(n(\omega_+)+1)+ \Gamma_-(n(\omega_-)+1)}{\Gamma_+ \theta(\omega_+) + \Gamma_- \theta(\omega_-)}.
\end{align}
For frequencies $v_d q > \omega$ and $\hbar(v_dq\pm \omega)\beta \gg 1$, the ratio above can be expanded as 
\begin{align}
    \frac{S_x(\omega,T)}{S_x(\omega)} \approx 1+e^{-\hbar v_d q \beta} \left( e^{-\hbar \omega \beta}- \frac{\Gamma_-}{\Gamma_+}e^{\hbar \omega \beta}
    \right)
\end{align}
showing that so long as $T_{AE} \gg T$, i.e., $\hbar v_d q \beta \gg 1$, finite temperature effects lead to a small perturbation of the zero-temperature power spectrum. 
\end{document}


\title{Supplemental Information for: Nonequilibrium thermodynamics of the acoustoelectric quantum vacuum}

\author{Ryan O. Behunin}
\email[]{ryan.behunin@nau.edu}
\affiliation{Department of Physics, Northern Arizona University, Flagstaff, AZ 86011, USA}
\affiliation{Center for Materials Interfaces in Research and Applications (<MIRA!), Flagstaff, AZ, USA}

\author{Andrew J. Shepherd}
\affiliation{Department of Physics, Northern Arizona University, Flagstaff, AZ 86011, USA}
\affiliation{Center for Materials Interfaces in Research and Applications (<MIRA!), Flagstaff, AZ, USA}

\author{Francesco Intravaia}
\affiliation{
Humboldt-Universit\"at zu Berlin, Institut f\"ur Physik, 12489 Berlin, Germany
}

\author{Matt Eichenfield}
\affiliation{Electrical, Computer, and Energy Engineering, University of Colorado Boulder and Sandia National Laboratories}

\maketitle

\section{Diagonalization of the acoustoelectric Hamiltonian}
\label{sec: diagonalization}
In this section, we show that the following operators
\begin{align}
    & x = \sum_j \int_{-\infty}^\infty d\omega' \ A_j(\omega') \hat{c}_j(\omega') = x^\dag
    \\
    &  p = -i\sum_j \int_{-\infty}^\infty d\omega' \ \omega'_{j} A_j(\omega') \hat{c}_j(\omega') = p^\dag
     \\
    & a_k(\omega) = \sum_j \int_{-\infty}^\infty d\omega' \ \chi_{kj}(\omega,\omega') \hat{c}_j(\omega') = \sum_j \int_{-\infty}^\infty d\omega' \ \chi_{k,-j}(\omega,-\omega') \hat{c}^\dag_j(\omega')
\end{align}
diagonalize the acoustoelectic Hamiltonian (Eqs. \eqref{eq: H_0} \& \eqref{eq: H_int} of the main text) where $\hbar \Omega b^\dag b = (p^2+\Omega^2 x^2)/2$. In addition to diagonalizing the Hamiltonian, these operators can also be shown to satisfy the equal-time commutation relations. The functions and operators above are defined by
\begin{align}
\label{eq: diag func 1}
    & A_j(\omega) = - \sqrt{2\hbar\Omega} g \chi_{\omega} \alpha(\omega_j) 
    \\
    & \chi_{kj}(\omega,\omega') = \delta_{kj} \delta(\omega-\omega') + \Delta\chi_{kj}(\omega,\omega') \\
    & \Delta\chi_{kj}(\omega,\omega') = \frac{2 \Omega g^2 \chi^*_\omega \chi_{\omega'} \alpha(\omega'_j)}{\omega_k-\omega_j'-i\varepsilon} = \frac{1}{\hbar \alpha^*(\omega_k)}\frac{ A^*_k(\omega) A_j(\omega') }{\omega_k-\omega_j'-i\varepsilon} = \frac{1}{\hbar}\frac{\sqrt{2 \hbar \Omega} g \chi^*_\omega A_j(\omega') }{\omega_k-\omega_j'-i\varepsilon}
    \\
\label{eq: diag func 4}
    & \hat{c}_j(\omega) = \hat{a}_j(\omega) \theta(\omega) + \hat{a}^\dag_{-j}(-\omega) \theta(-\omega).
\end{align}

Considering each term in the Hamiltonian, we obtain
\begin{align}
  &  \frac{1}{2}p^2 = \sum_{j,j'}\int d\omega d\omega' \frac{1}{2} \omega_j \omega_{j'}'A^*_j(\omega)A_{j'}(\omega')
    \hat{c}^\dag_j(\omega) \hat{c}_{j'}(\omega'),
    \\
 &   \frac{1}{2}\Omega^2 x^2 = \sum_{j,j'}\int d\omega d\omega' \frac{1}{2}\Omega^2 A^*_j(\omega)A_{j'}(\omega')
    \hat{c}^\dag_j(\omega) \hat{c}_{j'}(\omega'),
    \\
&    \sum_{k}\int_0^\infty d\nu \hbar \nu_k a_k^\dag(\nu)a_k(\nu) =
    \sum_{j,j'}\int d\omega d\omega'
    \sum_{k}\int_0^\infty d\nu \hbar \nu_k
    \chi^*_{kj}(\nu,\omega) \chi_{kj'}(\nu,\omega') \hat{c}^\dag_j(\omega) \hat{c}_{j'}(\omega'),
    \\
& \sum_{k}\int_0^\infty d\nu  \sqrt{2\hbar\Omega} g (\chi_{\nu}a_k(\nu)+\chi^*_{\nu}a^\dag_k(\nu))x =  \sum_{j,j'}\int d\omega d\omega' \sum_{k}\int_0^\infty d\nu  \sqrt{2\hbar\Omega} g \bigg(\chi_{\nu} \chi_{k,-j}(\nu,-\omega) 
\\
& \quad \quad \quad \quad \quad \quad \quad \quad \quad \quad \quad \quad \quad \quad \quad \quad \quad \quad \quad \quad 
+\chi^*_{\nu}\chi^*_{kj'}(\nu,\omega)
\bigg)A_{j'}(\omega') \hat{c}^\dag_j(\omega) \hat{c}_{j'}(\omega'). \nonumber 
\end{align}
Combining like terms, we express the Hamiltonian as $H = \sum_{j,j'} \int d\omega d\omega' I_{j,j'}(\omega,\omega') \hat{c}^\dag_j(\omega) \hat{c}_{j'}(\omega')$ where the integrand is given by
\begin{align}
    I_{j,j'}(\omega,\omega')
    = \frac{1}{2} (\omega_j \omega_{j'}'+ \Omega^2)A^*_j(\omega)A_{j'}(\omega')
   & +
    \sum_{k}\int_0^\infty d\nu \hbar \nu_k
    \chi^*_{kj}(\nu,\omega) \chi_{kj'}(\nu,\omega')
    \\
    &+
    \sum_{k}\int_0^\infty d\nu  \sqrt{2\hbar\Omega} g \bigg(\chi_{\nu} \chi_{k,-j}(\nu,-\omega)+\chi^*_{\nu}\chi^*_{kj}(\nu,\omega)
\bigg)A_{j'}(\omega'). \nonumber
\end{align}
Using the relation $\hat{c}_j(\omega) = \hat{c}^\dag_{-j}(-\omega)$, we are free to symmetrize the integrand according to 
\begin{align}
    I_{jj'}(\omega,\omega') \to I^{sym}_{jj'}(\omega,\omega') = \frac{1}{2} (I_{jj'}(\omega,\omega')+I_{-j',-j}(-\omega',-\omega) )
\end{align}
where a commutation of the operators neglects a constant contribution to the Hamiltonian that does not impact the dynamics. 

Next, we explicitly evaluate the integrals that contain a $\delta$-function
\begin{align}
    I_{j,j'}(\omega,\omega')
    = & \hbar \omega_j \delta_{jj'}\delta(\omega-\omega') \theta(\omega)+ \frac{1}{2} (\omega_j \omega_{j'}'+ \Omega^2)A^*_j(\omega)A_{j'}(\omega')
    +
    \hbar \omega'_{j'}\Delta\chi^*_{j'j}(\omega',\omega) \theta(\omega')
    \\
    & +
    \hbar \omega_{j}\Delta\chi_{jj'}(\omega,\omega') \theta(\omega)
    +
     \sqrt{2\hbar\Omega} g \bigg(\chi_{-\omega} \theta(-\omega)    
    +\chi^*_{\omega}\theta(\omega)
\bigg)A_{j'}(\omega')
     \nonumber
    \\
    &
    \sum_{k}\int_0^\infty d\nu \hbar \nu_k
    \Delta\chi^*_{kj}(\nu,\omega) \Delta\chi_{kj'}(\nu,\omega')
    +
    \sum_{k}\int_0^\infty d\nu  \sqrt{2\hbar\Omega} g \bigg(\chi_{\nu} \Delta\chi_{k,-j}(\nu,-\omega)+\chi^*_{\nu}\Delta\chi^*_{kj}(\nu,\omega)
\bigg)A_{j'}(\omega') \nonumber
\end{align}

After explicit symmetrization and simplification using Eqs. \eqref{eq: diag func 1}-\eqref{eq: diag func 4}, we find 
\begin{align}
    I^{sym}_{j,j'}(\omega,\omega') = & \delta{jj'} \delta(\omega-\omega')+\bigg[
    \frac{1}{2}\frac{\omega_j \tilde{\mu}(\omega_{j'}')-\omega_{j'}' \tilde{\mu}^*(\omega_{j})}{\omega_j -\omega_{j'}'-i\varepsilon}
    \\
&  + \sum_k \int_{-\infty}^\infty d\nu \frac{\nu_k(-\nu_k^2 +\omega_j^2
+\omega_j \omega_{j'}'+ {\omega_{j'}'}^2)}{(\nu_k^2-(\omega_j-i\varepsilon)^2)(\nu_k^2-(\omega_{j'}'+i\varepsilon)^2)} 2 \Omega g^2 |\chi_{\nu}|^2\bigg] A^*_j(\omega) A_{j'}(\omega')
\end{align}

Using the definition of the susceptibility from the main text (Eq. \eqref{eq: mu-susc}) we find
\begin{align}
    \frac{1}{2}\frac{\omega_j \tilde{\mu}(\omega_{j'}')-\omega_{j'}' \tilde{\mu}^*(\omega_{j})}{\omega_j -\omega_{j'}'-i\varepsilon}
    = & \sum_k \int_0^\infty d\nu \frac{2\Omega |g|^2\nu_k}{\omega_j-\omega_{j'}'-i \varepsilon}\bigg[\frac{\omega_j}{\nu_k^2-(\omega_{j'}'+i \varepsilon)^2}-\frac{\omega_{j'}'}{\nu_k^2-(\omega_{j}-i \varepsilon)^2}\bigg] |\chi_\nu|^2
    \\
    = & \sum_k \int_0^\infty d\nu \bigg[\frac{\nu_k(\nu_k^2-\omega_j^2-\omega_j \omega_{j'}' - {\omega_{j'}'}^2)}{(\nu_k^2-(\omega_{j'}'+i \varepsilon)^2)(\nu_k^2-(\omega_{j}-i \varepsilon)^2)}\bigg] 2\Omega |g|^2|\chi_\nu|^2.
\end{align}
Combining all of these results, we obtain
\begin{align}
  I^{sym}_{j,j'}(\omega,\omega') = & \delta{jj'} \delta(\omega-\omega'),  
\end{align}
proving the assertion that the Hamiltonian for this coupled system is 
\begin{align}
    H = \sum_j \int_0^\infty d\omega \hbar \omega_j \ \hat{a}^\dag_j(\omega) \hat{a}_j(\omega).
\end{align}

\bibliography{refs}